\documentclass[]{iopart}
\usepackage{iopams}
\usepackage{setstack}
\usepackage[]{graphicx}
\usepackage{color}
\usepackage[latin1]{inputenc}

\newcommand{\Prob}{{\rm Prob}}

\newcommand{\NA}{N_{\rm A}}
\newcommand{\rH}{{\rm H}}
\newcommand{\If}{{\rm If}}
\newcommand{\re}{{\rm e}}
\newcommand{\KJ}{K_{\rm J}}

\begin{document}

\title[]{Model selection in the average of inconsistent data: an analysis of the measured Planck-constant values}
\author{G Mana$^1$, E Massa$^1$ and M Predescu$^2$}
\address{$^1$INRIM -- Istituto Nazionale di Ricerca Metrologica, str.\ delle Cacce 91, 10135 Torino, Italy}
\address{$^2$Politecnico di Torino, corso Duca degli Abruzzi 24, 10129 Torino, Italy}

\begin{abstract}
When the data do not conform to the hypothesis of a known sampling-variance, the fitting of a constant to a set of measured values is a long debated problem. Given the data, fitting would require to find what measurand value is the most trustworthy. Bayesian inference is here reviewed, to assign probabilities to the possible measurand values. Different hypothesis about the data variance are tested by Bayesian model comparison. Eventually, model selection is exemplified in deriving an estimate of the Planck constant.
\end{abstract}

\submitto{Metrologia}


\ead{g.mana@inrim.it}

\section{Introduction}
Given a set of measured values of a constant and the associated uncertainties, the Gauss-Markov theorem states that the weighted arithmetic mean is the unbiased minimum-variance estimator of the measurand \cite{Luemberger}. The uncertainty of the mean, which is smaller than the smallest data uncertainty, does not depend on data spread. This is a consequence of the assumption that the variance of the sampling distribution of each datum is exactly know. In practice, this hypothesis is often false and the inconsistency of the data -- quantified, for example, by the $\chi^2$ or the Birge-ratio values -- suggests that the uncertainties associated with data are merely lower bounds to the standard deviations of their sampling distributions. In this case, the Gauss-Markov theorem is of no help and the choice of an optimal measurand value is a long debated issue. To average inconsistent data, a number of approaches have been considered; for an overview the reader can refer to \cite{Birge,Paule:Mandel,Weise:2000,Cox:2002,Kacker:2002,Willink:2002,Dose,Sivia} and the included references. Data inconsistency suggests many different hypotheses; for instance, that a term is missing in the error budget or that uncertainties are lower bounds to standard-deviations. Hence, it is necessary to identify the best data model.

Hypothesis testing is an important part of data analysis. In the averaging of inconsistent data, we are interested to quantify and to compare the degree of reliability of, for instance, Gaussian sampling having known variance, a missing term in the error budget, or sampling-variance underestimate. Traditional tests asses the mismatch between model and data; for instance, via $\chi^2$ calculation. Although a small model likelihood corresponds to a large $\chi^2$ value, to test a model from data, we must calculate the probability that the model is correct, not the misfit.

Decision theory, probability calculus, and Bayesian inference help to deal with the discrepancy between quoted uncertainties and data scatter. Given a data model, the first step is to find the post-data probability density of the measurand values, which is summarized by the product of data sampling-distribution and the pre-data distribution of the measurand values. The normalizing constant of this product, which is named evidence, measures of how much the data support the underlying model. Different models can thus be compared by the evidence calculation. Once the best model has been selected, the optimal choice of the measurand value minimizes the expected loss over the probabilities of the measurand values. Bayesian inference allows different hypotheses to be compared and makes no longer necessary to exclude the data disagreeing with the majority, as well as to arbitrarily scale the uncertainties to make the data consistent.

After review of data averaging, the assumption that data are sampled from independent Gaussian distributions having known variance is compared against alternative hypotheses, for example, that the uncertainties are only lower bounds to the standard deviations or that there are undiscovered errors. As an example, hypothesis testing is carried out by using the measured values of the Planck constant. The best data-model, of those considered, is identified and a Planck-constant value is estimated.

\section{Problem statement}
Let us consider a set of $N$ measured values $x_i$ of a measurand $h$, where $u_i$ is the uncertainty associated with the $x_i$ datum. The data are judged to be mutually inconsistent when combined in a least squares analysis. Hence, they are supposed independently sampled from Gaussian distributions having unknown variance $\sigma_i^2$. The problem is to find optimal estimates of the measurand value and of the confidence intervals.

We cannot rely on the weighted mean, because the actual variances of the sampling distributions are not known. On the other hand, we cannot go back to the arithmetic mean, because it leaves out significant information given by the data uncertainties.

Different hypotheses can be made about the  variance of data. The purpose is to choose between the hypotheses and to find an optimal value of the measurand. There is no constraint about hypotheses, nor there is on their number. As an illustration we shall consider: H0) the data uncertainties are the standard deviations of the sampling distributions; H1) the standard deviations are larger than the data uncertainties by a common scale factor \cite{Birge,Dose}, H2) the standard deviations are larger than the data uncertainties by a common additional contribution \cite{Paule:Mandel}; H3) the data uncertainties are lower bounds for the standard deviations of the relevant sampling distributions \cite{Sivia}. In the same way, we  can test the extend by which the data support any other model.

\section{Hypothesis testing}
Let us assume a basic knowledge of probability calculus and of Bayesian inference; introductions can be found in \cite{Sivia,Jaynes,McKay,Gregory}. To choose from different data models, we must compare the probabilities of each of them being true, given the $\{x_i\}$ data-set. To find the post-data probability $\Prob( \rH|x_i)$, where H indicates the hypothesis being tested, we start from the product rule
\begin{equation}\label{ProbAB}
 \Prob(A,B) = \Prob(A|B)\Prob(B) = \Prob(B|A)\Prob(A) ,
\end{equation}
where $A$ and $B$ are propositions -- e.g., affirming that a measurement result is $x_i$, that the measurand value is $h$, or that the hypothesis H is true -- and $A|B$ implies that $B$ is true. Hence,
\begin{equation}\label{ProbH}
 \Prob(\rH,x_i) = \Prob( \rH|x_i) W(x_i) = Z( x_i|\rH ) \Prob(\rH) ,
\end{equation}
where
\begin{equation}
 W(x_i) = \sum_\rH \Prob(\rH,x_i) =  \sum_\rH Z( x_i|\rH ) \Prob(\rH) ,
\end{equation}
is the probability density of data regardless of any hypothesis and $\Prob(\rH)$ is the probability of H regardless of data. The post-data probability of H,
\begin{equation}
 \Prob(\rH|x_i) = \frac{Z( x_i|\rH) \Prob(\rH)}{W(x_i)} ,
\end{equation}
is determined by two factors: $\Prob(\rH)$, the pre-data probability of H, and the data-dependent term $Z_\rH(x_i) = Z(x_i|\rH)$, which is named evidence for H. The $Z(x_i|\rH)$ evidence depends on both the data and H. For a fixed model, $Z(x_i|\rH)$ is the data distribution regardless of the model parameters. For fixed data, $Z(x_i|\rH)$ is the evidence for the data model. If we are only interested to compare H against another hypothesis, $W(x_i)$, which is a normalizing factor, can be ignored. In addition, on the assumption that, before the data are available, the probabilities of different hypotheses are the same, also $\Prob(\rH)$ can be ignored. Therefore, to find the most probable hypothesis, we can identify $Z(x_i|\rH)$ with $\Prob(\rH|x_i)$, the calculation of which is thus central.

The evidence calculation is based again the Bayesian analysis. To examine this point, let us consider the general case where $\rH=\rH(h,\lambda)$ is a class of hypotheses having $h$ and $\lambda$ as parameters. The post-data probability density $\Theta_\rH(h,\lambda;x_i)$ of the model parameters is obtained by application of
\begin{equation}\label{product}
 P(h,\lambda,x_i|\rH) = \Theta_\rH(h,\lambda;x_i) Z_\rH(x_i) = L_\rH(h,\lambda;x_i) \pi(h,\lambda) ,
\end{equation}
where $Z_\rH$ is the sought evidence for H, $\pi(h,\lambda)$ is the pre-data probability density of the parameter values, and $L_\rH(x_i; h,\lambda)$ is a data-dependent term. For fixed $h$ and $\lambda$, $L_\rH(x_i; h,\lambda)$ is the data sampling-distribution; for fixed data, $L_\rH(h,\lambda;x_i)$ is the likelihood of the parameter values. Here and in the following, in the function argument, the semicolon separates the independent variables from the fixed parameters.

Model selection within the H class is equivalent to an estimation problem. From (\ref{product}), for a fixed $\lambda$, the evidence for the $\rH(\lambda)$ model,
\begin{equation}\label{Z:l}
 Z_\rH(\lambda) = \int_{-\infty}^{+\infty} L_\rH(h,\lambda;x_i) \pi(h,\lambda)\, \rmd h
\end{equation}
is the marginalization of $L_\rH(h,\lambda;x_i) \pi(h,\lambda)$ with respect to $h$, that is, the non normalized post-data distribution of the $\lambda$ values, regardless of the measurand value. Similarly, the evidence for the whole H class,
\begin{equation}\label{Z:H}
 Z_\rH = \int_{-\infty}^{+\infty} L_\rH(h,\lambda) \pi(h,\lambda)\, \rmd h \, \rmd \lambda =
 \int_{-\infty}^{+\infty} Z_\rH(\lambda) \, \rmd \lambda ,
\end{equation}
is the normalizing factor of $L_\rH(h,\lambda;x_i) \pi(h,\lambda)$. If we only wish to choose an optimal measurand value, we can forget $Z_\rH $; however, in model selection problems, it is of the utmost importance.

\section{Measurand estimate}\label{problem}
Given the data and their associated uncertainty, the Bayesian solution to the estimate problem makes it necessary to assign probabilities to the $h$ values and to use these probabilities to minimize any given loss function \cite{Jaynes,McKay,Gregory}. The post-data distribution $\Theta_\rH(h;x_i)$ is central to this analysis. Given the loss $\mathfrak{L}(h_0-h)$ due to a wrong estimate $h_0$, the optimal choice of the $h$ value minimizes
\begin{equation}\label{cost}
 \langle \mathfrak{L}(h_0-h) \rangle = \int_{-\infty}^{+\infty} \mathfrak{L}(h_0-h) \Theta_\rH(h)\, \rmd h .
\end{equation}
With a quadratic loss, the optimal estimate is the mean; with a linear loss, it is the median. With a loss independent of error, it is the mode. Confidence intervals are easily calculable from $\Theta_\rH(h)$.

When H is a class of hypotheses, the result of the Bayesian analysis is the joint probability density $\Theta_\rH(h,\lambda; x_i)$, where $\lambda$ is the class parameter. In this case the usual approach to measurand prediction uses the probability density $\Theta_\rH(h,\lambda_0; x_i)$ with a fixed $\lambda_0$ value, e.g., the most probable $\lambda$ value. More accurate predictions are obtained by marginalization,
\begin{equation}
 \overline{\Theta}_\rH(h) = \int_{-\infty}^{+\infty} \Theta_\rH(h,\lambda)\, \rmd \lambda ,
\end{equation}
which takes account of the uncertainty of the $\lambda$ value.

\section{Model selection}
\subsection{{\rm H0} Hypothesis}
If the data uncertainties are the standard deviations, that is, if the data are independently sampled from Gaussian distributions having $\sigma_i^2=u_i^2$ variance, $L_\rH(h;x_i)$ is the Gaussian likelihood,
\begin{eqnarray}\label{Gauss}
 L_{\rH0}(h) &= \frac{1}{\sqrt{(2\pi)^N}\, \prod_{i=1}^N u_i}\, \exp \bigg[ - \sum_{i=1}^N \frac{(x_i-h)^2}{2u_i^2} \bigg]
 \\ \nonumber
 &= \frac{1}{\sqrt{(2\pi)^N}\, \prod_{i=1}^N u_i}\,
 \exp \bigg(- \frac{\chi^2}{2} \bigg)
 \exp \bigg[ - \frac{(\overline{x}-h)^2}{2\sigma_{\overline{x}}^2} \bigg] ,
\end{eqnarray}
where
\begin{numparts}\begin{equation}\label{x-mean}
 \overline{x} = \sigma_{\overline{x}}^2 \sum_{i=1}^N x_i/u_i^2
\end{equation}
is the weighted arithmetic mean of data,
\begin{equation}\label{s-mean}
 \sigma_{\overline{x}}^2 = \frac{1}{\sum_{i=1}^N 1/u_i^2}
\end{equation}
is the $\overline{x}$ variance, and
\begin{equation}
 \chi^2 = \sum_{i=1}^N x_i^2/u_i^2 -  \overline{x}^2 / \sigma_{\overline{x}}^2 =
 \sum_{i=1}^N \frac{(x_i-\overline{x})^2}{u_i^2}
\end{equation}\end{numparts}
is the sum of the squared residuals.

To calculate the post-data probability density of the $h$ value and the evidence for the H0 hypothesis, we must assign pre-data probabilities to the $h$ values. In the absence of any additional information, we suppose these pre-data probabilities independent of the unit-scale origin, which implies the uniform distribution
\begin{equation}\label{prior:h}
 \pi_h(h) = \frac{\If(\Xi/2 < h < \Xi/2)}{\Xi} ,
\end{equation}
where $\If(A)$ gives 1 if $A$ is true and 0 if it is not. The interval $[-\Xi/2,\Xi/2]$ includes the unknown measurand value. In the following, we shall always use this prior measurand distribution. Accordingly, the evidence for the H0 model is
\begin{equation}\fl\label{Z0}
 Z_{\rH0}  = \frac{1}{\Xi} \int_{-\Xi/2}^{+\Xi/2} L_{\rH0}(h)\, \rmd h =
           \frac{\sigma_{\overline{x}}}{\Xi \sqrt{(2\pi)^{N-1}}\, \prod_{i=1}^N u_i}\,
           \exp \bigg(- \frac{\chi^2}{2} \bigg) ,
\end{equation}
where $\Xi$ has been chosen large enough that the integration limits can be extended to infinity. The post-data probability density of the $h$ values is the Gaussian distribution
\begin{equation}\label{T0}
 \Theta_{\rH0}(h)  = \frac{1}{\sqrt{2\pi}\, \sigma_{\overline{x}}}\,
 \exp \bigg[ - \frac{(h-\overline{x})^2}{2\sigma_{\overline{x}}^2} \bigg] .
\end{equation}
The mean and most probable $h$ values are both equal to the weighted mean (\ref{x-mean}), whereas $\sigma_{\overline{x}}^2$ is the $h$ variance. In this case, the Bayesian analysis replicates the least-squares analysis. The novelty is the calculation of the evidence for the $\sigma_i=u_i$ hypothesis: it allows the probability of this assumption to be compared against alternatives.

\subsection{{\rm H1} Hypothesis}
The H1 hypothesis corresponds to $\sigma_i=\lambda u_i$, where $0 < \lambda < \Lambda$. Since H1 is a hypothesis class, we must specify the joint pre-data probability density $\pi(h,\lambda)=\pi_h(h)\pi_\lambda(\lambda)$ of both the $h$ and $\lambda$ parameters. This density summarizes the knowledge about $h$ and $\lambda$ before considering data. Unlike Dose -- who investigated extensively this case in \cite{Dose} -- in the absence of any information about the $\lambda$ magnitude prior to considering data, we use the bounded uniform density
\begin{equation}\label{prior:l}
 \pi_\lambda(\lambda) = \frac{\If(0 < \lambda < \Lambda)}{\Lambda} .
\end{equation}

Our choice is motivated by the H1 data-model including the $u_i$ values; therefore, the uncertainties associated to data are supposed available before the data themselves and fix the scale of the problem. For this reason, rather than $\pi_\lambda(\lambda)$ invariance with respect to scale transformations, namely, $\lambda = a\lambda'$ and $0 < \lambda' < \Lambda/a$, we suppose invariance with respect to translations, namely, $\lambda = \lambda_* + \lambda'$ and $-\lambda_* < \lambda' < \Lambda - \lambda_*$. The first invariance implies the Jeffryes prior distribution used in \cite{Dose,Sivia}; the second implies (\ref{prior:l}). We will use (\ref{prior:h}) and (\ref{prior:l}) throughout the paper.

The parameter likelihood,
\begin{numparts}\begin{equation}\fl\label{L:H1}
 L_{\rH1}(h,\lambda) = \frac{1}{\sqrt{(2\pi)^N} \lambda^N \prod_{i=1}^N u_i}\,
 \exp \bigg(- \frac{\chi^2}{2\lambda^2} \bigg)
 \exp \bigg[ - \frac{(\overline{x}-h)^2}{2\lambda^2\sigma_{\overline{x}}^2} \bigg] ,
\end{equation}
and the $\lambda$ evidence,
\begin{equation}\label{Z2}
 Z_{\rH1}(\lambda)  = \frac{\sigma_{\overline{x}}\, \If(0 < \lambda < \Lambda)}{\Xi \Lambda \sqrt{(2\pi)^{N-1}} \lambda^{N-1}
 \prod_{i=1}^N u_i}\, \exp \bigg(- \frac{\chi^2}{2\lambda^2} \bigg) ,
\end{equation}\end{numparts}
are obtained from (\ref{Gauss}) and (\ref{Z0}), where $\Xi$ has been chosen large enough that the integration limits can be extended to infinity.

Equation (\ref{Z2}) expresses the relative probability of different $\lambda$ values; the most probable one,
\begin{equation}\label{Birge}
 R_{\rm B} = \sqrt{\frac{\chi^2}{N-1}} ,
\end{equation}
is the Birge ratio. If $R_{\rm B}$ is not equal to 1, the data are inconsistent: the measurement uncertainties have been underestimated and the most probable standard deviations, $R_{\rm B} u_i$, differ from the uncertainties associated to data. The data are inconsistent also if $R_{\rm B} < 1$. In this case, the measurement uncertainties have been overestimated.

The Bayesian derivation of (\ref{Birge}) allows an expression of the Birge ratio to be found also when the data are correlated. By observing that the $\chi^2$ value is invariant with respect to orthogonal transformations of the $\bi{x}=[x_1, x_2, ...\,  x_N]^{\rm T}$ data set, if the data are correlated, the sum of the squared residuals is $\chi^2 = \bi{x}^{\rm T} \bi{C}_{xx}^{-1} \bi{x}$, where $\bi{C}_{xx}$ is the covariance matrix. Hence, in order to make data consistent, the covariance matrix must be scaled to $R_{\rm B} \bi{C}_{xx}$, where $R_{\rm B}$ is still given by (\ref{Birge}).

With a fixed $\lambda$, say, $\lambda=R_{\rm B}$, the post-data distribution of the measurand values is
\begin{equation}\label{H1:marginal1}
 \Theta_{\rH1}(h;R_{\rm B}) = N(h;\overline{x}, \sigma_B) = \frac{1}{\sqrt{2\pi}\, \sigma_B}\, \exp \bigg[ - \frac{(h-\overline{x})^2}{2\sigma_B^2} \bigg] ,
\end{equation}
where $\sigma_B=R_{\rm B} \sigma_{\overline{x}}$ and $N(h;\overline{x},\sigma_B)$ is a normal distribution having mean $\overline{x}$ and variance $\sigma_B^2$.

According to (\ref{Z:H}), the evidence for the H1 class is
\begin{equation}\label{Z:H1}
 Z_{\rH1} = \int_0^\Lambda Z_{\rH 1}(\lambda)\, \rmd \lambda =
 \frac{\sigma_{\overline{x}}\, \Gamma(N/2 - 1)} {2 \Lambda\Xi \sqrt{2\pi^{N-1}(\chi^2)^{N-2}} \prod_{i=1}^N u_i} ,
\end{equation}
where $\Lambda$ is large enough that the limit of integration can be extended to infinity and $N>2$. Consequently, by combining (\ref{L:H1}) and (\ref{Z:H1}), the joint post-data distribution of $h$ and $\lambda$ is
\begin{equation}\fl
 \Theta_{\rH1}(h,\lambda) = \frac{\sqrt{(\chi^2)^{N-2}}}
 { \sqrt{2^{N-3}\pi}\, \sigma_{\overline{x}}\, \Gamma(N/2 - 1)}
 \frac{ \displaystyle \exp \bigg(\frac{-\chi^2}{2\lambda^2} \bigg)
 \exp \bigg[ \frac{-(h-\overline{x})^2}{2\lambda^2\sigma_{\overline{x}}^2} \bigg] }
 {\lambda^N} ,
\end{equation}
where $\Xi, \Lambda \rightarrow \infty$ and $N>2$. The marginal distribution of the measurand regardless of $\lambda$ is then
\begin{equation}\label{H1:marginal2}
 \overline{\Theta}_{\rH1}(h) = \frac{ \sqrt{(\chi^2)^{N-2}}\,\Gamma\big[(N-1)/2\big] }
 { \sqrt{\pi}\, \sigma_{\overline{x}}\, \Gamma(N/2 - 1)}
 \left[ \frac{1}{\chi^2 + (h-\overline{x})^2/\sigma_{\overline{x}}^2} \right]^\frac{N-1}{2} ,
\end{equation}
where the space of the $h$ values has been extended to infinity and $N>2$.

The availability of analytic results is of great help because it allows the different ways for estimating the measurand value and the confidence intervals to be compared. The first way relies on distribution (\ref{H1:marginal1}). In this case, the $h$ mean is $\overline{x}$ and the 68\% confidence interval is $[\overline{x}-\sigma_B,\overline{x}+\sigma_B]$. This approach assumes that the standard deviations of data are $R_{\rm B} u_i$ with certainty, but $R_{\rm B} u_i$ are only the most probable values. The second way solves this ambiguity. It relies on (\ref{H1:marginal2}) and accounts of the lack of knowledge about the standard-deviation of data. In this case, though the $h$ mean is still $\overline{x}$, the 68\% confidence interval must be calculated numerically from the cumulative distribution function of (\ref{H1:marginal2}).

Another and last consideration concerns the presence in evidence (\ref{Z:H1}) of the ranges $\Xi$ and $\Lambda$ of the $h$ and $\lambda$ pre-data distributions. They are Occam's factors penalizing H1 for having the free parameters $h$ and $\lambda$ \cite{McKay}. The larger are these ranges, i.e., the model freedom to explain data, the smaller is the evidence for H1 given by data. Since we shall consider the same pre-data distributions, these Occam's factors are of no relevance and will be omitted when comparing the H1, H2, and H3 models.

\subsection{H2 Hypothesis}
The H2 class corresponds to $\sigma_i^2=u_i^2+\lambda^2 u_o^2$, where $0 < \lambda < \Lambda$, $u_0$ is is a reference $u$ value, and the pre-data distributions of $h$ and $\lambda$ are (\ref{prior:h}) and (\ref{prior:l}). The parameter likelihood, model evidences, and post-data probability densities are obtained from (\ref{Gauss}), (\ref{Z0}), and (\ref{T0}), where $u_i^2+\lambda^2 u_0^2$ substitutes for $u_i^2$. It is not possible to give them a concise form, but this does not prevent numerical computations.

\subsection{{\rm H3} Hypothesis}
We now suppose that the data uncertainties are lower bounds for the standard deviations, that is, $u_i \le \sigma_i \le \max(u_i, \lambda u_0)$, where $0 < \lambda < \Lambda$ and $u_0$ is a reference $u$ value. A variant is $u_i \le \sigma_i \le \max(u_i, \lambda u_i)$. In both cases, the pre-data distributions of $h$ and $\Lambda$ are (\ref{prior:h}) and (\ref{prior:l}). Subsequently, we must assign probabilities to the unknown $\sigma_i$ values, for which only the $u_i$ lower bounds are known.

In the absence of any information about the $\sigma_i$ magnitude before considering the data, but with the uncertainties $u_i$ being already known, we use the translation invariant uniform distributions
\begin{numparts}\begin{eqnarray}\label{prior}
 P(\sigma;u,\lambda) = \frac{{\rm If}(u<\sigma<\lambda u_0)}{\lambda u_0 -u} \qquad \lambda u_0 > u ,
\end{eqnarray}
if the model is $u\le \sigma \le \max(u, \lambda u_0)$, and
\begin{eqnarray}
 P(\sigma;u,\lambda) = \frac{{\rm If}(u<\sigma<\lambda u)}{(\lambda-1)u}  \qquad \lambda > 1 ,
\end{eqnarray}\end{numparts}
if the model is $u \le \sigma \le \max(u, \lambda u)$; the $i$ subscript has been dropped.

$P(\sigma;u,\lambda)$ being stated, the sampling distribution of each $x$ datum can be marginalized to eliminate the unknown variance. Hence, if $u\le \sigma \le \max(u, \lambda u_0)$,
\begin{numparts}\begin{eqnarray}
 \fl Q(x;h,u,\lambda) = \int_u^{\lambda u_0} N(x;h,\sigma)P(\sigma;u,\lambda)\, \rmd \sigma \nonumber \\ \label{Q1}
 = \frac{\Gamma\big[0,(x - h)^2/(2\lambda^2 u_0^2)\big] - \Gamma\big[0,(x - h)^2/(2u^2)\big]}
 {2\sqrt{2\pi} (\lambda u_0-u)} \qquad & \lambda u_0 > u\nonumber \\
 = N(x;h,u) & \lambda u_0 \le u
\end{eqnarray}
where $\Gamma(a,z)$ is the incomplete Euler gamma function. If $u \le \sigma \le \max(u, \lambda u)$, the marginalized sampling distribution is
\begin{eqnarray}
 \fl Q(x;h,u,\lambda) = \frac{\Gamma\big[0,(x - h)^2/(2\lambda^2 u^2)\big] - \Gamma\big[0,(x - h)^2/(2u^2)\big]}
 {2\sqrt{2\pi} (\lambda-1)u} \qquad & \lambda > 1 \nonumber \\
 =  N(x;h,u) & \lambda \le 1 .
\end{eqnarray}\end{numparts}
For fixed $\lambda$ values, the post-data probability density of the measurand is
\begin{numparts}\begin{equation}\label{TH3}
 \Theta_{\rH3}(h; \lambda) = \frac{\pi_h(h)\pi_\lambda(\lambda)L_{\rH3}(h;\lambda)}{Z_{\rH3}(\lambda)} ,
\end{equation}
where
\begin{equation}\label{L1}
 L_{\rH3}(h;\lambda) = \prod_{i=1}^N  Q(x_i;h,u_i,\lambda),
\end{equation}
$Q$ is given in (\ref{Q1}-$b$), $x_i$ and $u_i$ are the $i$-th datum and its uncertainty, and the evidence for the $\rH3(\lambda)$ model is
\begin{equation}\label{Z1}
 Z_{\rH3}(\lambda) = \pi_\lambda(\lambda) \int_{-\infty}^{+\infty} \pi_h(h) L_{\rH3}(h,\lambda)\, \rmd h .
\end{equation}\end{numparts}
It is not possible to obtain an analytical form of (\ref{Z1}), which must be evaluated numerically.

\begin{table}
\caption{Values of the Planck constant. The main reference is \cite{CODATA}; when relevant, more specific references are listed in the table. The values determined from the $h/m({\rm n})$, $h/m(\re)$, $h/m({\rm Cs})$, $h/m({\rm Rb})$, $h/\Delta m(^{29}{\rm Si})$, and $h/\Delta m(^{33}{\rm S})$ ratios depend on the same $\NA$ value.}
\begin{center}\begin{tabular}{lllr}
\hline
method &laboratory &$10^{34} h$ / J s  &reference\\
\hline
\multicolumn{4}{c}{measurement of the Faraday constant} \\
 $h/e$       &NIST 1980                  &$6.6260657(88)  $  &\cite{Bower:1980} \\
\hline
\multicolumn{4}{c}{$\NA$ and $\gamma_{\rm lo}$ measurement} \\
 $h/m(\re)$  &NIST 1989, IAC 2011                  &$6.62607122(73) $  &\cite{Andreas,Williams:1989} \\
 $h/m(\re)$  &NIM 1995, IAC 2011                   &$6.6260686(44)  $  &\cite{Andreas,Liu:1995} \\
 $h/m(\re)$  &KRISS/VNIIM 1998, IAC 2011           &$6.6260715(12)  $  &\cite{Andreas,Park:1999} \\
\hline
\multicolumn{4}{c}{$\gamma_{\rm hi}$ measurement} \\
 $h/e$   &NPL 1979                      &$6.6260729(67) $  &\cite{Kibble:1979} \\
 $h/e$   &NIM 1995                      &$6.626071(11)  $  &\cite{Liu:1995} \\
\hline
\multicolumn{4}{c}{$\KJ$ measurement} \\
 $h/(2e)$   &NMI 1989                   &$6.6260684(36) $  &\cite{Clothier:1989} \\
 $h/(2e)$   &PTB 1991                   &$6.6260670(42) $  &\cite{Funck:1991} \\
\hline
\multicolumn{4}{c}{watt-balance experiments} \\
 $h/m({\mathfrak{K}})$ &NPL 1990      &$6.6260682(13) $  &\cite{Kibble:1990} \\
 $h/m({\mathfrak{K}})$ &NIST 1998     &$6.62606891(58)$  &\cite{Williams:1998} \\
 $h/m({\mathfrak{K}})$ &NIST 2007     &$6.62606901(34)$  &\cite{Steiner:2007} \\
 $h/m({\mathfrak{K}})$ &METAS 2011    &$6.6260691(20) $  &\cite{METAS:2011} \\
 $h/m({\mathfrak{K}})$ &NPL 2012      &$6.6260712(13) $  &\cite{NPL:2012} \\
 $h/m({\mathfrak{K}})$ &NRC 2012      &$6.62607063(43)$  &\cite{NRC:2012} \\
\hline
\multicolumn{4}{c}{$\NA$ and quotient of $h$ and the neutron mass} \\
 $h/m({\rm n})$ &PTB 1998, IAC 2011   &$6.62606887(52)$ &\cite{Andreas,Krueger:1998,Krueger:1999} \\
\hline
\multicolumn{4}{c}{$\NA$ and quotient of $h$ and the mass of the electron or an atom} \\
 $h/m(\re)$      &CODATA 2006, IAC 2011   &$6.62607003(20) $ &\cite{CODATA,Andreas} \\
 $h/m({\rm Cs})$ &Stanford 2002, IAC 2011 &$6.62607000(22) $ &\cite{Andreas,Wicht:2002} \\
 $h/m({\rm Rb})$ &LKB 2011, IAC 2011      &$6.62607009(20) $ &\cite{Andreas,Bouchendira:2011} \\
 weighted mean    &                       &$6.62607005(21) $ \\
\hline
\multicolumn{4}{c}{$\NA$ and quotient of $h$ and nuclear mass defects} \\
 $h/\Delta m(^{29}{\rm Si})$ &MIT/ILL 2005, IAC 2011     &$6.6260764(53) $ &\cite{Andreas,Rainville:2005} \\
 $h/\Delta m(^{33}{\rm S})$  &MIT/ILL 2005, IAC 2011     &$6.6260686(34) $ &\cite{Andreas,Rainville:2005} \\
weighted mean                 &                           &$6.6260709(29) $ \\
\hline
\end{tabular} \label{planck-values-table} \end{center}
{\small NIST: National Institute of Standards and Technology (USA), IAC: International Avogadro Coordination, NIM: National Institute of Metrology (People's Republic of China), NPL: National Physical Laboratory (UK), NMI: National Metrology Institute (Australia), PTB: Physikalisch Technische Bundesanstalt (Germany), METAS: Federal Office of Metrology (Switzerland), NRC: National Research Council (Canada), Stanford: Stanford University (USA), LKB: Laboratoire Kastler-Brossel (France), MIT: Massachusetts Institute of Technology (USA), ILL: Institut Laue-Langevin (France)}
\end{table}

\section{Planck constant estimate}
\subsection{Input data}
As an example of application, let us consider the choice of a Planck constant value on the basis of the measurement results listed in table \ref{planck-values-table} \cite{CODATA,CODATA:2010}. The Planck constant links energy and momentum to the frequency and the wavelength of the wave-function. Therefore, the $h$ determinations match energy (momentum) and frequency (wavelength) measurements. According to whether an electric or mechanic system is considered, the measurement result is a value of the $h/e$ or $h/m$ ratio, where $e$ and $m$ are the electron charge and a mass.

The electric determinations are based on the measurements of $h/e$ or $h/(2e)$ by Ag $\rightarrow$ Ag$^+ + e^-$ electrolysis, the high-field spin-flip frequency, $\gamma_{\rm hi}$, of protons in samples of H$_2$O, and the tunneling of Cooper's pairs in Josephson junctions. The watt-balance experiments accede directly to the $h/m_\mathfrak{K}$ ratio, where $m_\mathfrak{K}$ is the mass of the international kilogram prototype. The measurement of the Avogadro constant $N_A$ opened the way towards comparing the watt-balance measurements of $h$ with the $h/m$ quotients, where $m$ is mass of a particle or of an atom. The $h/m(n)$ ratio was determined by time-of-flight measurements of monochromatic neutrons; low-field magnetic resonance, optical spectroscopy and atom interferometry were instrumental to determine $h/m({\rm e})$, $h/m({\rm Cs})$, and $h/m({\rm Rb})$. Nuclear spectroscopy of $^{28}$Si and $^{32}$S, after capture of a thermal neutron, allowed the $\NA h$ product to be determined by comparing the mass defect of $^{29}{\rm Si}$ and $^{33}{\rm S}$ with the frequencies of the $\gamma$ photons emitted in the cascades from the capture state to the ground state.

\begin{figure}
\centering
\includegraphics[width=125mm]{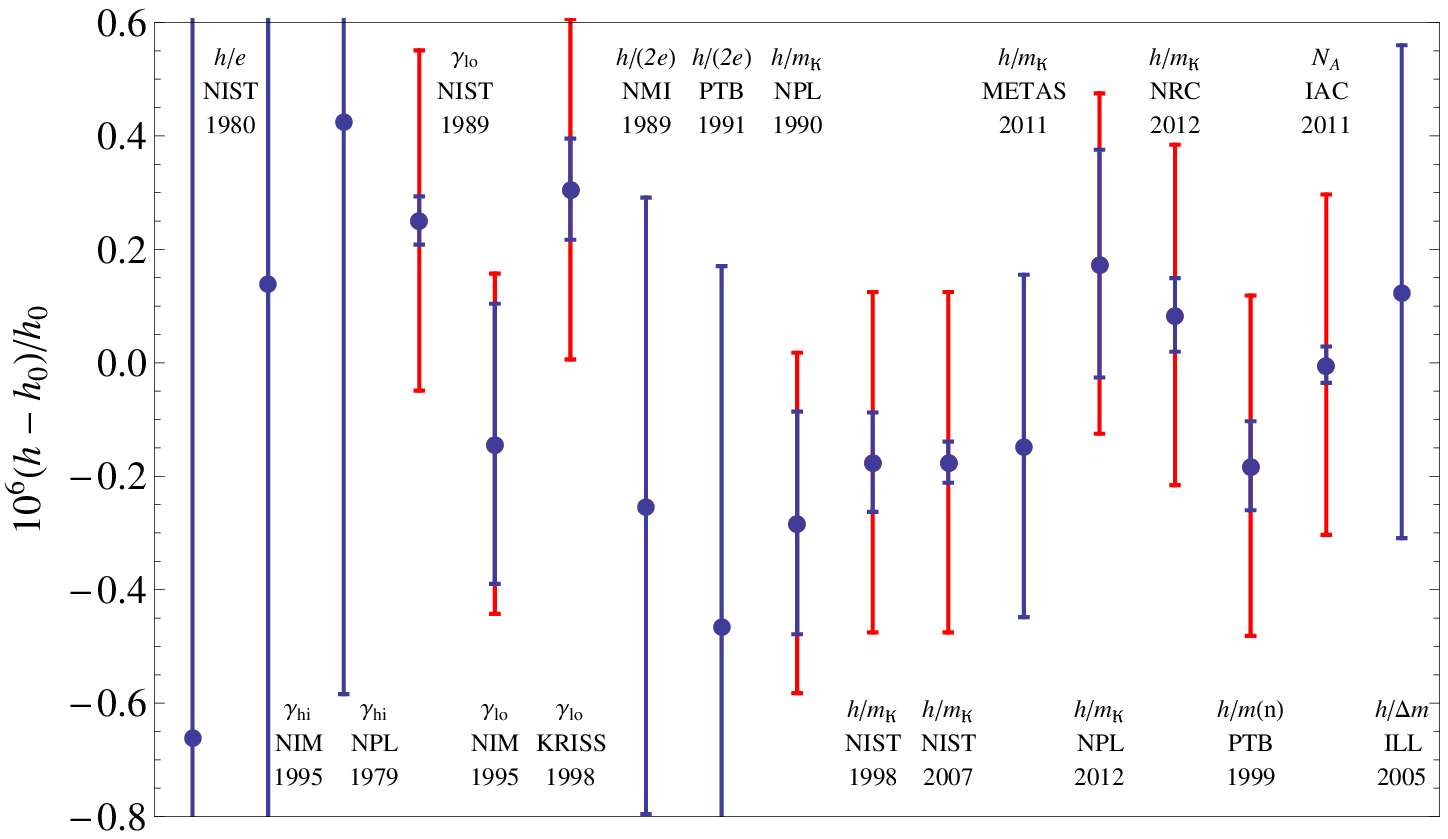}
\caption{Values of the Planck constant given in table \ref{planck-values-table}. The label $\NA$ indicates the weighted mean of values determined from the $h/m(\re)$, $h/m({\rm Cs})$, $h/m({\rm Rb})$ ratios. The label $h/\Delta m$ indicates the weighted mean of values determined from the $h/\Delta m(^{29}{\rm Si})$, and $h/\Delta m(^{33}{\rm S})$ ratios. The reference value is the weighted mean, $h_0=6.62607007\times 10^{-34}$ J s. The double error bars (blue and red) indicate the lower and upper bounds to the standard deviation, according to the most probable data model.}\label{planck-values-plot}
\end{figure}

Figure \ref{planck-values-plot} shows the input data used in the analysis. The values calculated from the measurements of $h/m({\rm e})$, $h/m({\rm Cs})$, and $h/m({\rm Rb})$ have been averaged, because their uncertainty is mainly affected by the $N_A$ determination and, therefore, they are highly correlated. The values obtained from nuclear spectroscopy have been averaged because they were obtained by repetitions of the same experiment. For the sake of numerical simplicity, in the subsequent analyses, we set $u_0=10^{-6}h_0$ and use the reduced data $10^6(x_i-h_0)/h_0$ and uncertainties $10^6u_i/h_0$, where $h_0=6.62607007\times 10^{-34}$ J~s is the weighted arithmetic mean of data.

\subsection{{\rm H0} hypothesis}
When supposing the data sampled from independent Gaussian distributions the standard deviations of which are the data uncertainties, the most probable $h$ value is the weighted arithmetic mean. The evidence for this hypothesis is $Z_{\rH0} = 9.6 \times 10^{-16} \Lambda \mathfrak{V}$. The awful numerical value of $Z_{\rH0}$ is due to fact that it is the normalizing factor of $L_{\rH0}(h;x_i) \pi_h(h)$. Its dimensions are (J~s)$^{-N}$ and the $\mathfrak{V} = 10^{6(N-1)}/(\Lambda \Xi h_0^{N-1})$ coefficient originates from the normalizations of the data and uncertainties used in the calculation.

\subsection{{\rm H1} Hypothesis}
\begin{figure}
\centering
\includegraphics[width=63.5mm]{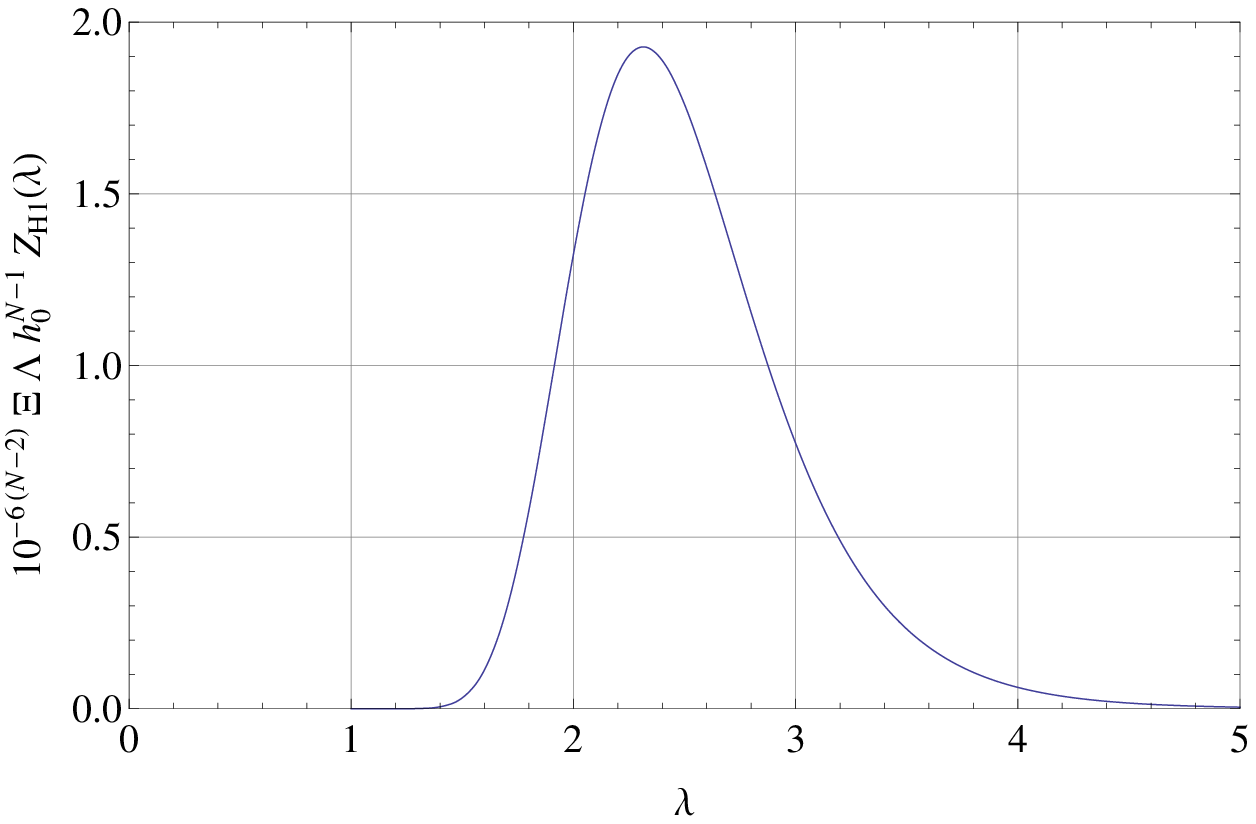}
\includegraphics[width=62.5mm]{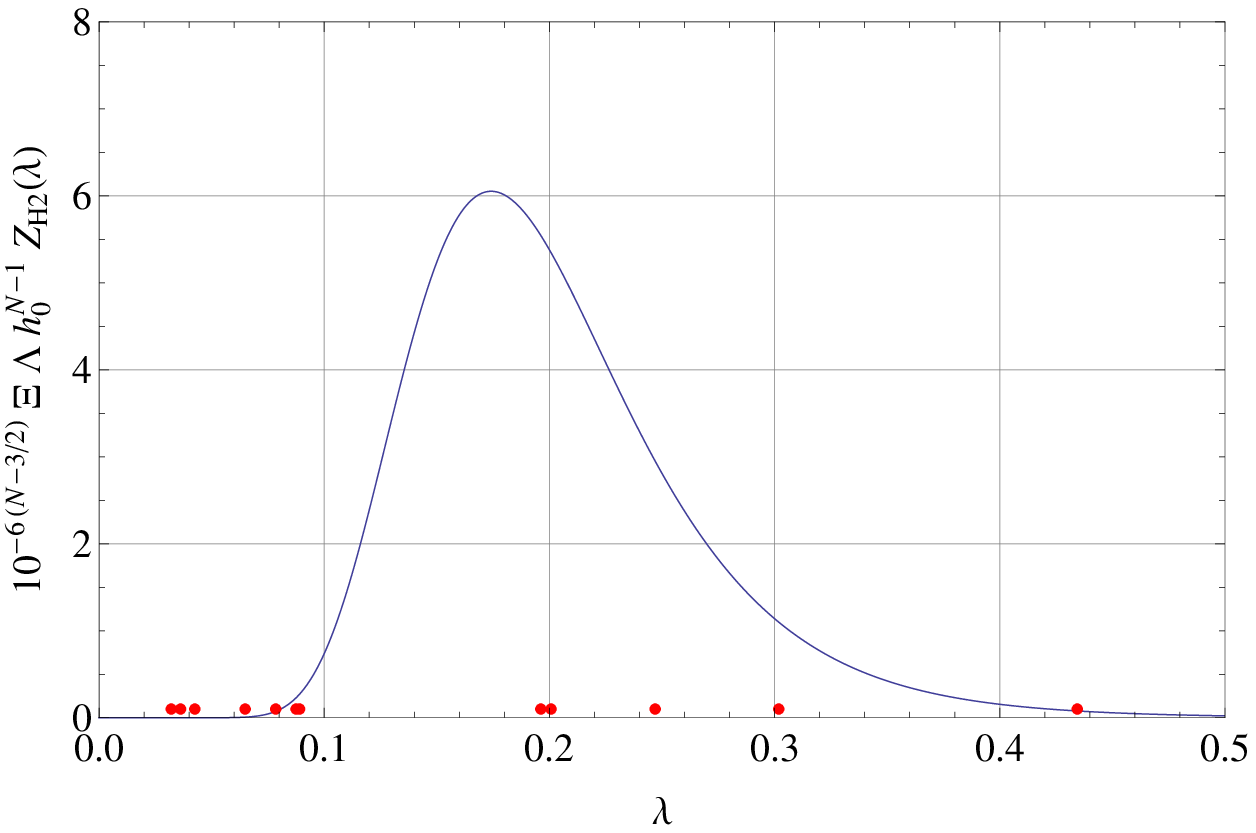}
\includegraphics[width=63.5mm]{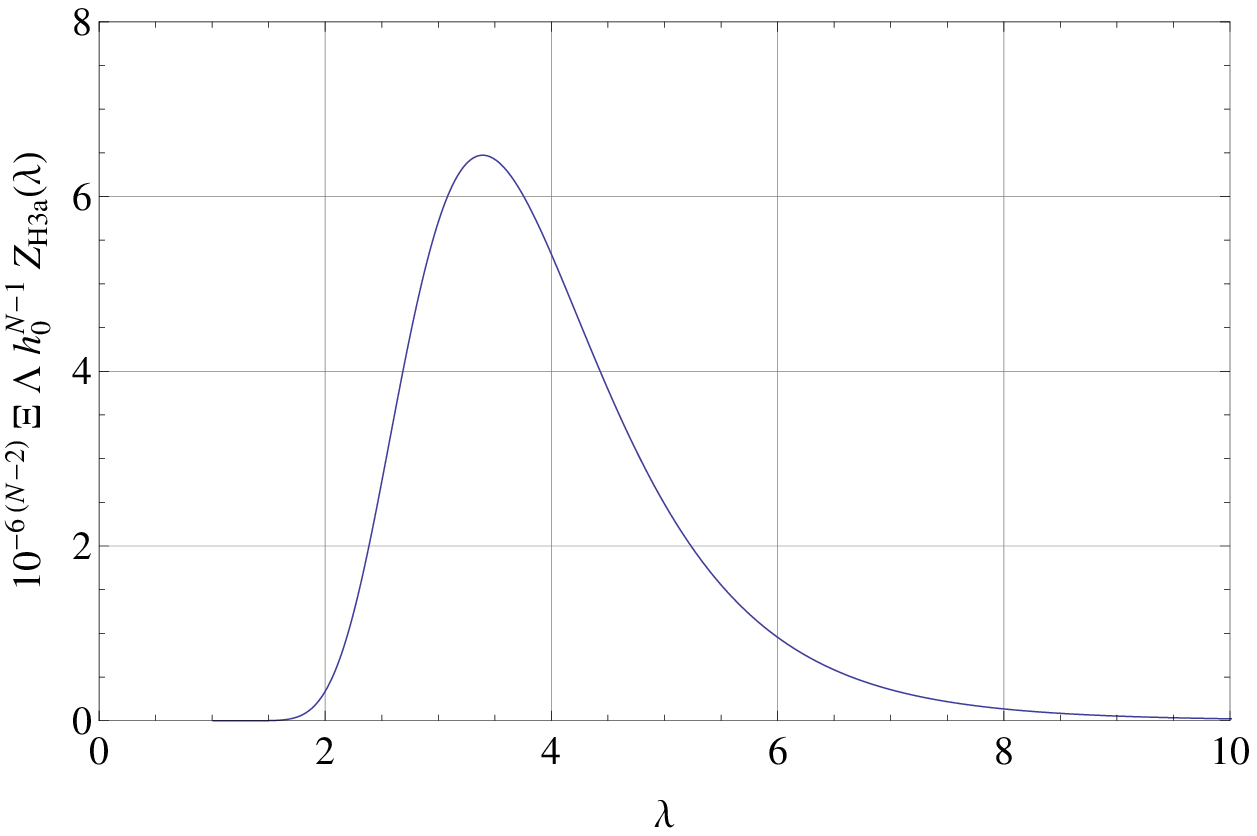}
\includegraphics[width=62.5mm]{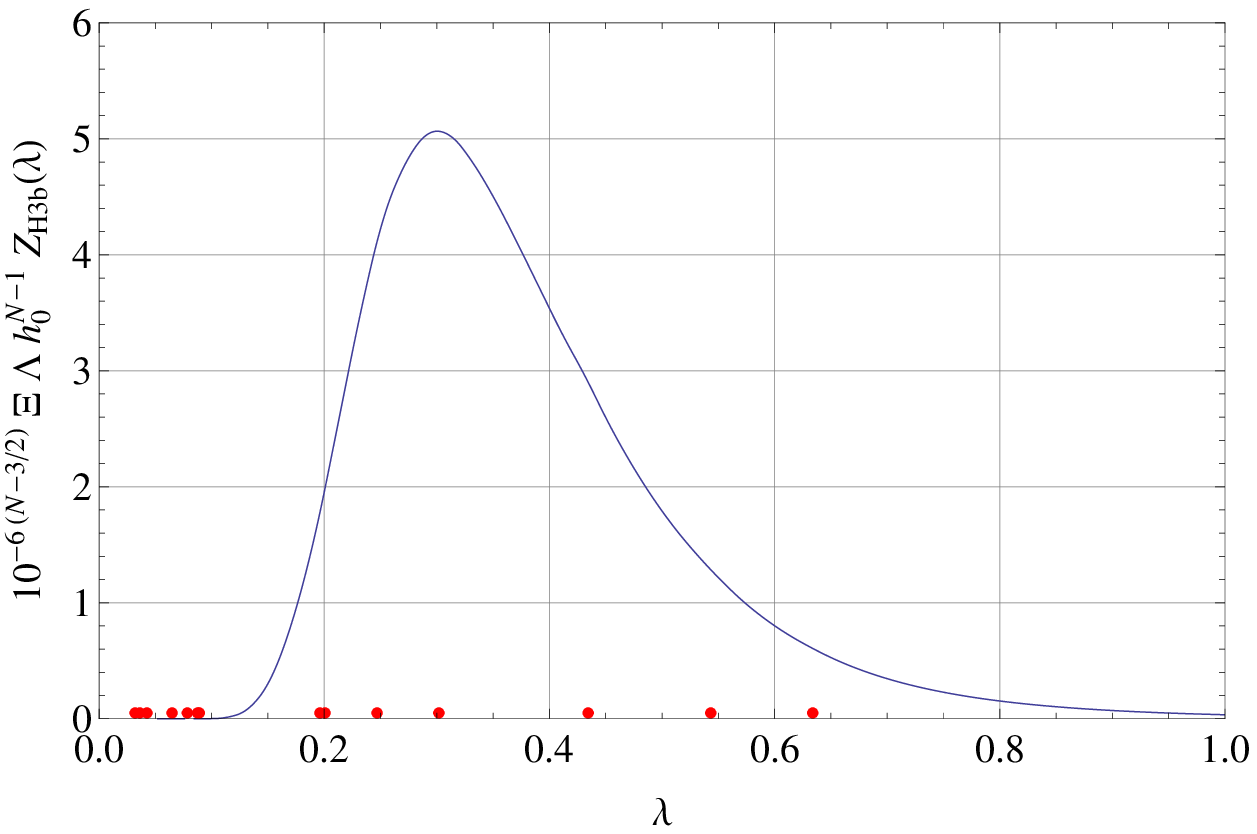}
\centering
\caption{Evidences for the $\sigma = \lambda u$ (top, left), $\sigma^2 = u^2 + (10^{-6}\lambda h_0)^2$ (top, right, $u < \sigma < \max(u, \lambda u)$ (bottom, left), and $u < \sigma < \max(u,10^{-6}\lambda h_0)$ (bottom, right) hypotheses. The red dots indicate the uncertainties associated to the data.}\label{H1:fig}
\end{figure}

Though none of the data is totally out of scale, with $\chi^2 = 85.69$ for $\nu=N-1=16$ degrees of freedom and Birge ratio $R_{\rm B}=\sqrt{\chi^2/\nu} = 2.31$, they are nevertheless inconsistent. In order to make the data consistent, we supposed that the quoted uncertainties are wrong by a common scale factor $\lambda$. The evidence for this hypothesis is shown in Fig.\ \ref{H1:fig}; the most probable $\lambda$ value is the Birge ratio. By multiplying the data uncertainties by $R_{\rm B}$, the $\chi^2$ of the weighted arithmetic mean is 16 and the Birge ratio is 1; this removes the discrepancy between data spread and uncertainties.

The evidence for the whole of H1 is $Z_{\rH1} = 2.1 \times 10^{-6}\mathfrak{V}$. Without fixing the extension $\Lambda$ of the parameter space, we cannot compare the H0 and H1 hypotheses, but the $\rH1(\lambda=1)$ model replicates the H0 hypothesis. Hence, the comparison of the $Z_{\rH1}(\lambda=1) = 9.6 \times 10^{-16}\mathfrak{V}$ evidence against $Z_{\rH1}(\lambda=R_{\rm B}) =1.9\times 10^{-6}\mathfrak{V}$ confirms that the odds are severely against the hypothesis of a Gaussian samplings having known variance.

\subsection{{\rm H2} Hypothesis}
This hypothesis conjectures a missing uncertainty-contribution $10^{-6}\lambda h_0$ common to all the data. The evidence for this hypothesis is shown in Fig.\ \ref{H1:fig}. The most probable $\lambda$ value is $\lambda_0 = 0.17$, with $Z_{\rH2}(\lambda_0) = 6.1\times 10^{-3}\mathfrak{V}$. The evidence for the whole of H2 is $Z_{\rH2} = 8.0\times 10^{-4}\mathfrak{V}$. The comparison against the H0 hypothesis can be made by observing that $\rH2(\lambda=0)$ replicates H0 and that $Z_{\rH2}(\lambda=0) = 9.6 \times 10^{-16}\mathfrak{V}$. As before, the odds are against the hypothesis of a Gaussian samplings having known variance.

\subsection{{\rm H3} Hypothesis}
When the data uncertainties are supposed to be the lower bounds of unknown standard deviations, we compared two hypotheses. In the H3a case, the standard deviation upper-bounds are proportional to the uncertainty associated with each datum, i.e., $u_i \le \sigma_i \le \max(u_i, \lambda u_i)$. In the H3b case, the same $10^{-6}\lambda h_0$ value bounds the greatest standard-deviation of each datum, i.e., $u_i \le \sigma_i \le \max(u_i,10^{-6}\lambda h_0)$. The evidences for these hypotheses are shown in Fig.\ \ref{H1:fig}. Also in these cases, ${\rm H3a}(\lambda=1)$ and ${\rm H3b}[\lambda=\min(10^6 u_i/h_0)]$ replicate H0.

The H3 hypothesis could be thought pessimistic: the uncertainties $u_i$ set only lower bounds for the standard deviations of the sampling distributions, which are supposed very large if at all possible. However, as shown in Fig.\ \ref{H1:fig}, an infinite standard deviation is not supported by the data and finite upper bounds exist. In the H3a class, the optimal upper bound is $4.3 u_i$, with $Z_{\rm H3a}(4.3 u_i) = 6.5 \times 10^{-6}\mathfrak{V}$. In the H3b class, the optimal upper bound is $0.30\times 10^{-6}h_0$, with $Z_{\rm H3b}(0.30\times 10^{-6}h_0) = 5.1\times 10^{-3}\mathfrak{V}$. This eliminates the worry of a too pessimistic view. The evidences for the whole of these hypothesis classes are given in table \ref{h:estimates}.

\subsection{Prediction of the Planck constant value}
\begin{figure}
\centering
\includegraphics[width=75mm]{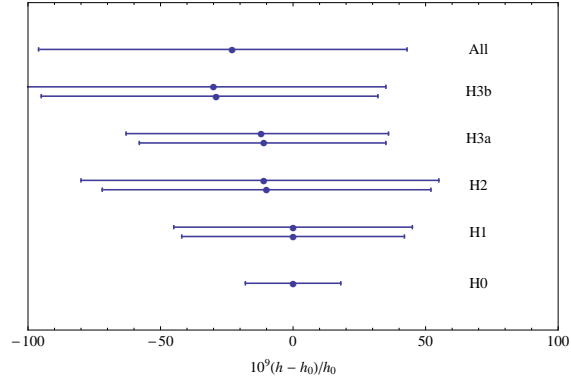}
\caption{Predictions of the Planck constant values derived from the marginal distributions $\Theta_\rH(h,\lambda_0)$ (lower line) and $\overline{\Theta}_\rH(h)$ (upper line). The last value (All) takes into account the ignorance of the data model by weighing the H1, H2, H3a, and H3b values according to the relevant evidences. The reference $h$ value is the weighted mean, $h_0=6.62607007 \times 10^{-34}$ J s of the data. The asymmetric error bars indicate 68\% confidence intervals.}\label{results}
\end{figure}

Predictions of the $h$ value were made from both the marginal distributions $\Theta_\rH(h;\lambda_0)$, where only the most probable $\lambda=\lambda_0$ value was taken into account, and $\overline{\Theta}_\rH(h)$, where all the possible $\lambda$ values have been considered. The median $h$ value and 68\% confidence intervals are given in table \ref{h:estimates} and Fig.\ \ref{results}, together with the evidences for the different models. The weighted arithmetic mean of the data, which is also the median of the marginal distributions based on the H0 and H1 models, has been taken as the reference value.

\begin{table}
\caption{Medians $\overline{h}$ of the Planck constant values calculated according to the marginal distributions $\Theta_\rH(h,\lambda_0)$ and $\overline{\Theta}_\rH(h)$, where $\lambda_0$ is the most probable value of the model parameter. The reference $h$ value is the weighted mean, $h_0=6.62607007 \times 10^{-34}$ J s of the data. The lower and upper bounds are the 16\% and 84\% quantiles. $Z_\rH$ is the evidence for the model, which is proportional (through the same proportionality factor) to the probability $\Prob(\rH|x_i)$ of being true.}
\begin{center}\begin{tabular}{llrrll}
\hline \\
model &$\lambda_0$
&$\displaystyle\frac{10^9(\overline{h}-h_0)}{h_0}$
&$\displaystyle\frac{10^9(\overline{h}-h_0)}{h_0}$
&$\displaystyle\frac{\Xi \Lambda h_0^{N-1} Z_\rH}{10^{6(N-1)}}$
&$\Prob(\rH|x_i)$ \vspace{1mm} \\
                       &                    &\multicolumn{2}{c}{distribution used} \\
                       &                    &$\Theta_\rH(h,\lambda_0)$        &$\overline{\Theta}_\rH(h)$ \vspace{1mm} \\
\hline \\
H0    &$-$                      &$-$                     &$\;\;\;\;0_{-18}^{+18}$ &$9.6\times 10^{-16}\Lambda$ &$-$ \vspace{1mm}\\
H1    &2.3                      &$\;\;\;\,0_{-42}^{+42}$ &$\;\;\;\;0_{-45}^{+45}$ &$2.1\times 10^{-6}$
&$0.95\times 10^{-4}$ \vspace{1mm}\\
H2    &$0.17$  &$-10_{-62}^{+62}$       &$-11_{-69}^{+66}$       &$0.8\times 10^{-3}$  &0.36 \vspace{1mm}\\
H3a   &3.4                      &$-11_{-47}^{+46}$       &$-12_{-51}^{+48}$       &$1.5\times 10^{-5}$
&$6.82\times 10^{-3}$ \vspace{1mm}\\
H3b   &$0.30$  &$-29_{-66}^{+61}$       &$-30_{-74}^{+65}$       &$1.4\times 10^{-3}$  &0.63 \vspace{1mm}\\
all   &$-$                      &$-$                     &$-23_{-73}^{+66}$       &$2.2\times 10^{-3}$  &1.00 \vspace{1mm}\\
\hline
\end{tabular} \label{h:estimates} \end{center}
{\small H0: $\sigma=u$, H1: $\sigma=\lambda u$, H2: $\sigma^2=u^2+(10^{-6}\lambda h_0)^2$, H3a: $u<\sigma<\max(u,\lambda u)$, H3b: $u<\sigma<\max(u,10^{-6} \lambda h_0)$}
\end{table}

To choose the best data model, we must compare the probabilities of each of them being true. These probabilities are proportional to the evidences given in table \ref{h:estimates}. Of those considered, the hypothesis most favoured by data is H3b. It postulates that the data uncertainties are lower bounds to the standard deviations of the sampling distributions and that a common upper bound exists for all of them. The most probable upper bound, $0.30\times 10^{-6}h_0$, is relatively small. Only some data have smaller uncertainty; in Fig.\ \ref{planck-values-plot} they are indicated by the double error bars. The evidence for hypothesis H2, which postulates that a common contribution is missing in the uncertainty budged, is almost high as the evidence for H3b. Besides, H2 and H3b differ only in the greater freedom allowed by H3b to the magnitude of the missed contribution.

\begin{figure}[b]
\centering
\includegraphics[width=75mm]{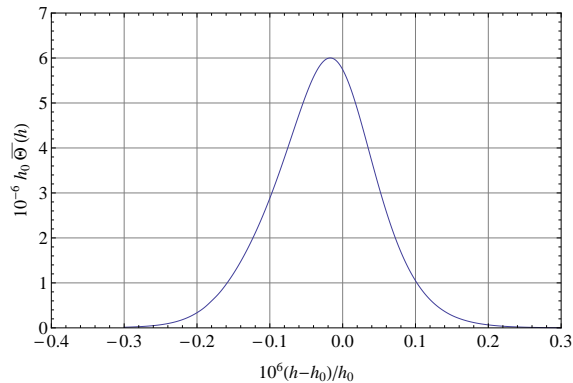}
\caption{Post-data distribution of the Planck constant values given the measurement results listed in table \ref{planck-values-table}. Marginalization has been carried out with respect to all models H1, H2, H3a, and H3b.}\label{post:data}
\end{figure}

Birge's hypothesis of a missing scaling factor common to all uncertainties, both in the original and extended versions H1 and H3a, though more supported than the hypothesis of Gaussian samplings having standard deviations equal to the data uncertainties, is not sustained by the data.

At the end of this analysis we have different estimates related to the different hypotheses about the uncertainties associated to data. The variability of the data model can be eliminated by marginalization. Hence,
\begin{equation}
 \overline{\Theta}(h) = \sum_\rH \overline{\Theta}_\rH(h)\Prob(\rH|x_i) = \frac{ \sum_\rH \overline{\Theta}_\rH(h) Z_\rH} {\sum_\rH Z_\rH} ,
\end{equation}
where (\ref{ProbH}) has been used and the same 25\% probability -- regardless of data -- has been supposed for each hypothesis. The numerical values of the relevant evidence and the final $h$ estimate are given in table \ref{h:estimates} and Fig. \ref{results}. The posterior density $\overline{\Theta}(h)$, which accounts for all the data models here considered, is shown in Fig.\ \ref{post:data}.

\section{Conclusions}
The probability calculus offers a rigorous solution to the problem of fitting a constant to a set of measured values, no matter whether the data are consistent or not. When we suspect the presence of outliers \cite{Sivia} or data non-conformity \cite{Weise:2000}, the post-data distribution of the measurand values depends on the model supposed for the unknown variance of the data.

Different data models can be considered and that most supported by data can be identified by calculating and comparing the probabilities of each model being true. Once the most probable model is found, a measurand value can be optimally chosen according the measurand post-data probabilities. The ignorance about the data model can also be taken into account by weighing the measurand estimates according to the model probabilities. This corresponds to marginalization of the post-data distribution of the measurand values over the data models. Our analysis clarifies that a hypothesis cannot be accepted or rejected absolutely, but only relatively to alternatives against which its probability is compared. From this viewpoint, hypothesis test and model selection are nothing else than a Bayesian inference procedure and they do not differ from parameter estimate.

Our analysis proved also that, on the assumption that the uncertainties associated to data are in error by an unknown scale factor, the Birge ratio can be interpreted as the most probable value of such scale factor. In addition, the analysis suggested how the Birge ratio can be generalized to the case when the data are correlated.

\ack
This work was jointly funded by the European Metrology Research Programme (EMRP) participating countries within the European Association of National Metrology Institutes (EURAMET) and the European Union.

\end{document}